# A method for testing bullets at reduced velocity

Michael Courtney, PhD and Amy Courtney, PhD
Ballistics Testing Group, P.O. Box 24, West Point, NY 10996
Michael_Courtney@alum.mit.edu, Amy_Courtney@post.harvard.edu

**Abstract:** Reconstruction of shooting events occasionally requires testing of bullets at velocities significantly below the typical muzzle velocity of cartridge arms. Trajectory, drag, and terminal performance depend strongly on velocity, and realistic results require accurately reconstructing the velocity. A method is presented for testing bullets at reduced velocities by loading the bullet into a sabot and firing from a muzzle loading rifle with a suitably reduced powder charge. Powder charges can be safely reduced to any desirable level when shooting saboted bullets from a muzzleloader; in contrast, cartridge arms can only be safely operated within a narrow window of powder charges/muzzle velocities. This technique is applicable to a wide range of both pistol and rifle bullets at velocities from 700 ft/s to 2000 ft/s.

## I. Introduction

The forensic reconstruction of shooting events occasionally requires testing of bullets at velocities significantly below the typical muzzle velocity of the firearm of interest. Similarly, understanding the drag and terminal performance of sporting arms at long ranges also requires testing of bullets at reduced velocities. Terminal performance parameters such as energy transfer, temporary cavitation, bullet expansion and penetration depth all depend, in part, on impact velocity [FAC96, MAC94, WWZ04, MAS92, COC07a]. Data also show that the ballistic coefficient, an important bullet parameter characterizing air drag, changes with velocity [COC07b].

Most cartridge arms can only be loaded safely over a narrow range of muzzle velocities [NOS96, HOR91, SPE94, BAR97]. For example, the Hornady Handbook of Cartridge Reloading [HOR91] lists a range of velocities from 1025 ft/s to 1150 ft/s for 124 grain bullets in the 9mm Parabellum and a range of velocities from 2400 ft/s to 2800 ft/s in the .308 Winchester. A thorough search of the literature will occasionally reveal a load that is tested to be safe at somewhat reduced velocities, but it is rare to find loads more than 10% or so slower than the slowest loads revealed by consulting two or three reloading manuals.

Several factors allow muzzleloaders to safely fire saboted bullets at any powder charge below the maximums listed by the manufacturers (typically 100-150 grains). These include the greatly reduced bore friction of a plastic sabot compared with a copper jacketed bullet, the non-progressive burn rate of black powder and approved muzzle loading substitutes compared with smokeless powder, the lower gas volumes of black powder and approved muzzle loading substitutes, the design of muzzle loading arms which handle gases safely in an unsealed breech, and the lower operating pressures compared with cartridge arms.

These factors combine to give assurance that the maximum barrel pressure in a muzzleloading rifle using black powder or an acceptable substitute will be a monotonic function of the powder charge for a given combination of sabot and bullet. In other words, decreasing the powder charge below the maximum will always decrease the peak barrel pressure. In contrast, in cartridge arms using smokeless powder, decreasing the powder charge below loads that have been tested safe by reliable sources can create unexpected pressure spikes which exceed the pressure rating of the firearm, thus endangering the shooter.

## II. Method

Muzzleloaders are capable of shooting saboted pistol bullets in a range of calibers. This fact has been used to test handgun bullets in deer [COC07c]. A .45 caliber in-line muzzleloader can accurately shoot saboted .355, .357, and .40 caliber pistol bullets over a wide range of velocities (700-2200 fps). A .50 caliber in-line muzzleloader can accurately shoot saboted .40, .429, .451, and .458 caliber bullets over a similar velocity range. Plastic sabots are commercially available[1] to shoot bullets of these diameters, and sabots designed for .357 caliber pistol bullets can reliably fire .358 caliber rifle bullets also, as long as the twist rate of the muzzleloader is adequate to stabilize the bullet. Figure 1 shows a number of bullets and two shotgun slugs in commercially available sabots.

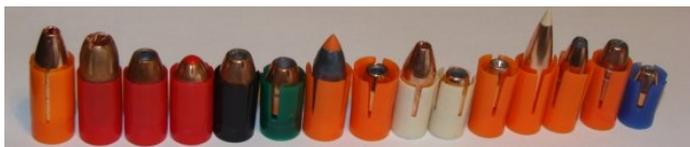

Figure 1: A variety of projectiles loaded in sabots for firing from .50 caliber (8 leftmost) and .45 caliber (7 rightmost) muzzleloaders. Projectiles include a 300 grain .458 caliber rifle bullet (1st from left), 20 gauge shotgun slugs removed from original sabots (2nd and 3rd from left), .429 caliber 44 magnum/special bullets (5th and 6th from left), .40 caliber .40 S&W/10mm bullets (8th and 10th from left), 9mm handgun bullets (1st and 5th from right), and .358 caliber rifle bullets (3rd and 4th from right).

Bullets with diameters incompatible with commercially available plastic sabots can be fired in sabots of the next larger size by taking up the extra space by wrapping the bullet base in a cotton or paper patch to assure a snug fit between the bullet and sabot prior to loading. This technique is illustrated in Figure 2.

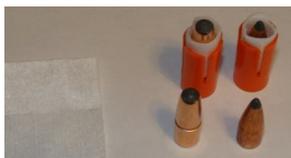

Figure 2: Two .308 caliber rifle bullets (150 grain on left, 110 grain on right) shown with a cotton cleaning patch used to assure snug fit between the .308" outer bullet diameter and the .357" inner diameter of the orange plastic sabot.

It is also possible to paper patch larger projectiles without the use of a plastic sabot. This technique was used with good accuracy from the Civil War era until the smokeless powder era began in the early 20th century. The space between an undersize projectile and the larger bore diameter is taken up with a paper patch that assures a snug fit. In principle, this technique could be used to fire projectiles too large to fit into commercially available plastic sabots.

Engraving of rifling can affect a bullet's ballistic coefficient [COC07b], and it is also possible that rifling marks can affect terminal performance by providing pre-stressed areas in the bullet jacket prior to impact.[2] Forensic ballistics experts have well-established techniques for recovering engraved bullets that are otherwise undeformed. Full-metal jacket bullets can often be recovered undeformed (other than rifling marks) from a water tank. Dependable recovery of expanding bullet designs requires shooting the bullet into a medium which decelerates the bullet with minimal force over a sufficiently large distance. Bullets that are undeformed (except for rifling) can also be recovered (occasionally, but not reliably) from the soft earth or decaying logs behind the target area at outdoor shooting ranges. The method presented allows for testing of reduced velocity drag and terminal ballistics of both previously unfired bullets (without rifling marks) and suitably recovered previously fired bullets (with rifling marks).

Different muzzleloaders produce widely varying muzzle velocities with comparable projectile weights and powder charges. Therefore, any bullet being tested requires a carefully developed powder charge to achieve the desired velocity in the specific muzzle loading rifle being used. For this reason, publishing a comprehensive list of reduced loads and resulting velocities is neither desirable nor useful. However, it is instructive to present some limited data to show the relevant trends, even though performance in different muzzle loading rifles might be much different.

---

[1] Many well-stocked shooter supply sources have a good selection of sabots. We have used [KNI08] and [PRB08].

[2] A number of bullet designs assure adequate performance by pre-stressing the bullet jacket and/or core material. To our knowledge, there is no published data regarding whether the stress of engraving the rifling has any impact on terminal performance.

Table 1 shows the resulting average measured muzzle velocity from a .50 caliber muzzleloader (TC Encore) shooting a 250 grain saboted bullet with the black powder substitute Hodgdon Triple Seven FFFG. Muzzle velocities were measured with an Oehler 35 chronograph with proof channel. Powder charge, though expressed in grains, is actually measured by an equivalent volume.

| Powder Charge (by volume) | Average Muzzle Velocity |
| --- | --- |
| 10 grains | 678 feet per second |
| 20 grains | 1025 feet per second |
| 30 grains | 1288 feet per second |

Table 1: Muzzle velocity vs. powder charge for a .50 caliber muzzleloader firing a 250 grain saboted bullet with reduced charges of Triple Seven FFFg powder.

Table 2 shows the resulting average measured muzzle velocity from a .45 caliber muzzleloader (custom Hubbard barrel on a New England Firearms frame) shooting a 115 grain bullet with Triple Seven FFFG. Muzzle velocities were measured with an Oehler 35 chronograph with proof channel. Powder charge is measured by an equivalent volume.

Triple Seven is a much more energetic powder than black powder or Pyrodex. Black powder or Pyrodex might be preferred for low velocity loads. In early 2009, we will be testing the potential to use powders based on a mixture of sugar and potassium nitrate.[3]

---

[3] This combination has often been suggested as a black powder substitute, and a blend (by weight) of 65% $KNO_3$ and 35% sucrose (table sugar) is widely used instead of black powder in model rocketry. It is slightly less energetic than black powder, and has a lower dependence of burn rate on pressure, suggesting smooth, gradual changes of muzzle velocity with powder charge. We plan to test a mixture of 65% $KNO_3$ and 35% sucrose blended in a coffee grinder, and compare performance with the addition of 2% red iron oxide as a catalyst. We expect smoother performance at lower charges, easier cleanup, and less foul smell than black powder and substitutes. Interested readers may contact us for preliminary results.

| Powder Charge (by volume) | Average Muzzle Velocity |
| --- | --- |
| 30 grains | 1150 feet per second |
| 40 grains | 1400 feet per second |
| 50 grains | 1530 feet per second |
| 60 grains | 1560 feet per second |
| 70 grains | 1750 feet per second |
| 80 grains | 1870 feet per second |

Table 2: Muzzle velocity vs. powder charge for a .45 caliber muzzleloader firing a 115 grain saboted bullet with reduced charges of Triple Seven FFFg powder.

### III.        Conclusion/Discussion

Both terminal performance and ballistic coefficient vary significantly with velocity. Testing bullet performance at lower velocity presents a challenge because smokeless powder charges cannot be safely lowered to arbitrary levels in cartridge arms, and extending the range to very large distances (so that air resistance slows the bullet to the desired level) is impractical because the required accuracy is probably not available. In other words, at the range where the velocity has slowed to the desired level, it is unlikely that the test will be able to reliably hit the tissue simulant (for studying terminal performance) or the chronograph window (for measuring velocity loss to infer ballistic coefficient).

This paper presents a method for shooting a wide variety of rifle and pistol bullets (and several shotgun slugs) at velocities much lower than available in cartridge arms using smokeless powder. Bullets are inserted into a suitable plastic sabot and fired from a muzzle loading rifle at reduced velocity. This technique can be combined with established techniques for measuring ballistic coefficient [COC07d], measuring bullet velocity [COU08], gelatin testing [WOL91], and studying terminal performance in living targets [COC07c].

However, there are several caveats. A load needs to be developed to achieve the desired velocity. Not every projectile can be tested at reduced velocity. One needs a combination of sabot and patching to match the diameter of bullet and barrel, and one needs a suitable rate of rifling twist to stabilize the bullet. Finally, if rifling marks can possibly affect the performance parameters being tested, valid testing requires first firing a bullet from a cartridge arm into a medium where the bullet can be recovered without further deformation to produce a projectile with rifling marks that can subsequently be saboted and fired from a muzzleloader at reduced velocity.

In spite of these caveats, the approach of using reduced muzzleloader charges compares quite favorably with the alternative of using unrifled pneumatic devices [WWZ04, MTY07], which are usually unrifled and unable to fire bullets which maintaining proper orientation in flight.

## REFERENCES/BIBLIOGRAPHY:

## About the Authors

*Amy Courtney* earned a MS in Biomedical Engineering from Harvard University and a PhD in Medical Engineering and Medical Physics from a joint Harvard/MIT program. She has taught Anatomy and Physiology as well as Physics. She has served as a research scientist at the Cleveland Clinic and Western Carolina University, as well as on the Biomedical Engineering faculty of The Ohio State University and the Physics faculty of the United States Military Academy at West Point.

*Michael Courtney* earned a PhD in experimental Physics from the Massachusetts Institute of Technology. He has served as the Director of the Forensic Science Program at Western Carolina University and also been a Physics Professor, teaching Physics, Statistics, and Forensic Science. Michael and his wife, Amy, founded the Ballistics Testing Group in 2001 to study incapacitation ballistics and the reconstruction of shooting events. www.ballisticstestinggroup.org